\documentclass[12pt,a4paper]{article}
\usepackage{epsfig}
\begin{document}
\title{Nucleon resonances and processes involving strange particles}
\author{S. Ceci, A. \v{S}varc and B. Zauner\\
Rudjer Bo\v{s}kovi\'{c} Institute, \\
Bijeni\v{c}ka c. 54, \\ 
10 000 Zagreb, Croatia\\ 
E-mail: Alfred.Svarc@irb.hr}

\maketitle

\begin{abstract}
An existing single resonance model with S11, P11 and P13 Breit-Wiegner resonances in the s-channel has been 
re-applied to the old \mbox{$\pi$N $\rightarrow$ K$ \Lambda$} data. It has been shown that the standard set of 
resonant parameters fails to reproduce the shape of the differential cross section. The resonance parameter 
determination has been repeated  retaining the most recent knowledge about the nucleon resonances.  The extracted set 
of parameters has confirmed the need for the strong contribution of a  P11(1710) resonance. The need for any 
significant contribution of the P13 resonance has been eliminated. Assuming that  the Baker. et al data 
set\cite{Bak78} is a most reliable one, the P11 resonance can not but be quite narrow. It emerges as a good candidate 
for the non-strange counter partner of the established pentaquark anti-decuplet.
\end{abstract}

In spite of the fact that the experimental data for the process \mbox{$\pi$N $\rightarrow$ K$ \Lambda$}, which show a 
distinct peeking  around the energy range of 1700 MeV, are available for quite some time\cite{Bak78,Dat78} the 
existence of  a P$_{11}$(1710) resonance in that energy range is not generally accepted, but even 
questioned\cite{Arn85,Arn04}. However, all coupled channel models\cite{PWA} do accept the P$_{11}$(1710) state as a 
legal and needed state, and the general agreement is that it is strongly inelastic. 

The confirmation  and the additional proof for the existence of the P$_{11}$(1710) resonance turned out to be 
critically needed quite recently in light of reported observations of exotic pentaquark $\Theta$(1539) and 
$\Xi$(1862) states\cite{Pen04}, as the mentioned state turns out to belong quite naturally into the pentaquark 
anti-decuplet configuration predicted by recent chiral soliton\cite{Dia97} and $q^4 \bar{q}$ strong color-spin 
correlated\cite{Jaf03} models.  

Keeping in mind the accumulated knowledge about nucleon resonances\cite{PWA} we have applied the existing, single 
resonance model of $\pi$N $\rightarrow$ K$ \Lambda$ process\cite{Tsu00} to the existing data set\cite{Dat78}, and 
allowed for the explicit presence of a narrow P$_{11}$ resonance. The single resonance model calculation has been 
repeated  using the standard set of parameters for the  S11(1650), P11(1710) and P13(1720) resonances of 
ref.\cite{PDG},  and they are given in Table 1 denoted as "PDG".  As it is shown in Figs.1 and 2. (thin solid line) 
that choice of parameters reproduces only the absolute value of the total cross section quite well, and manages to 
reproduce the shape of the angular distribution {\bf only} at w=1683 MeV. It fails miserably in reproducing the shape 
of the differential cross section at other energies. To eliminate the problem we have fitted the  $K \Lambda$ 
branching ratios in the Breit-Wigner parameterization  of the afore discussed resonances to the available 
experimental data set, enforcing a good description of the absolute value of the total cross section {\bf 
simultaneously}  keeping  the  shape of the {\em differential} cross sections of the \mbox{ $\pi$N $\rightarrow$ K$ 
\Lambda$} process linear in $cos(\theta)$ (indicated by experimental data of ref.\cite{Bak78}). \\
The following resonance parameter extraction method has been applied: \\
we have started with the belief that when a set of resonance parameters (masses, widths and branching fractions) is 
once established in any analysis using channels other then $K \Lambda$ channel, the only parameter which we are 
allowed to vary is a {\em branching fraction} to $K  \Lambda$ channel, {\em  while everything else (masses, widths, 
branching fractions to other channels) can not be changed}.  Hence, as  starting values we have used the resonant 
parameters for the $S_{11}$, $P_{11}$ and $P_{13}$ resonances obtained in the coupled channel analysis of 
$\pi$-nucleon scattering based on the   $\pi$-elastic and \mbox{$\pi$N $\rightarrow \eta$ N} channels\cite{Bat98}. 
The only parameter which was allowed to vary was the branching fraction to the $K \Lambda$ channel. 

\begin{table}[!h]
\begin{center}
Resonance parameters for the single resonance model.
{\footnotesize
\begin{tabular}{@{}|c|cccc|@{}}
\hline
{} &{} &{} &{} &{} \\[-1.5ex]
{} & {$M$}&$\Gamma[MeV]$ &$x_{\pi N}$[\%]& $x_{K \Lambda}$[\%]\\
{} & $S_{11}$ {} $P_{11}$ {} $P_{13}$&$S_{11}$ {} $P_{11}$ {} $P_{13}$&$S_{11}$ {} $P_{11}$ {} $P_{13}$&$S_{11}$ {} 
$P_{11}$ {} $P_{13}$\\[1ex]
\hline \hline 
{} &{} &{} &{} &{} \\[-1.5ex]
$PDG$ &1650 {}   1710 {}  1720 &150   {} 100  {}   150 &70   {}  15   {}  15   &7    {}  15  {}   6.5  \\[1ex]
$Sol  \ 1$ &1652  {} 1713  {}  1720 &202  {}   180 {}  244& 79 {}  22  {}  18 &{\em 2.4}   {}  {\em 23} {}  {\em 
0.16}\\[1ex]
$Sol  \ 2$ &1652 {}  1713  {}  1720 &202  {}   180  {}  244& 79 {}  22  {}  18 & {\em 2.4}  {}   {\em 35}  {}  {\em 
0.16}\\[1ex]
$Sol  \ 3$ &1652 {}  {\bf 1700} {}   1720 &202  {}   {\bf 60}  {}  244& 79 {}  22 {}   18 & {\bf 4}  {}   {\bf 30}  
{} {\em 0.16}\\[1ex]
\hline
\end{tabular}\label{table2} }
\vspace*{-13pt}
\end{center}
\end{table}
\noindent
  
\begin{figure}[!h] 
\begin{center}
\epsfig{figure=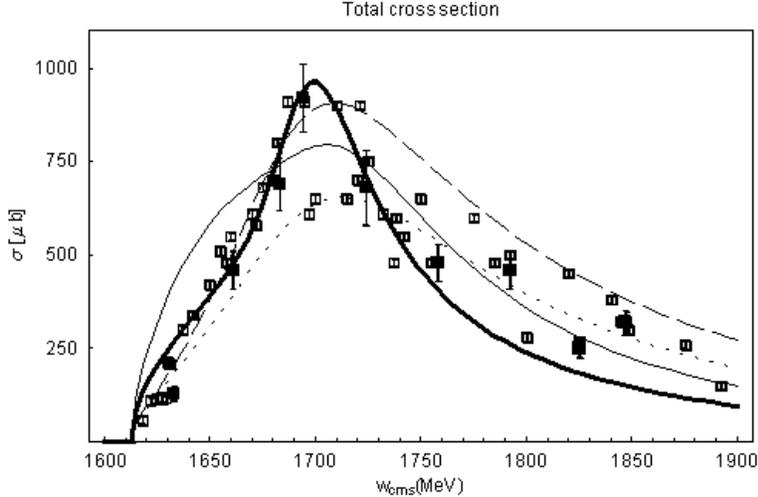,width=10cm}   
\caption{The agreement of the available experimental data for the total cross section (ref.\protect\cite{Bak78} (full 
boxes); ref.\protect\cite{Dat78}-open boxes) with the single resonance model predictions using different inputs for 
the resonance parameters: "standard" (PDG) set (thin solid line); $Sol \ 1$ (dotted line); $Sol \ 2$ (dashed line) 
and $Sol \ 3$ (thick solid line). \label{inter}}
\end{center}
\end{figure}
The three solutions for the choice of resonant parameters are obtained, and are together with the "standard" (PDG) 
solution given in Table 1. The agreement with the experimental data is given in Figs. 1. and 2. 
In extracting new resonance parameters we have kept in mind that the overall data set\cite{Bak78,Dat78} is mutually 
inconsistent. In addition, as the latest measured set of data\cite{Bak78} shows a surprisingly narrow width when 
compared to the overall trend, we have treated it separately with special care.

\begin{figure}[!h]
\begin{center} 
\epsfig{figure=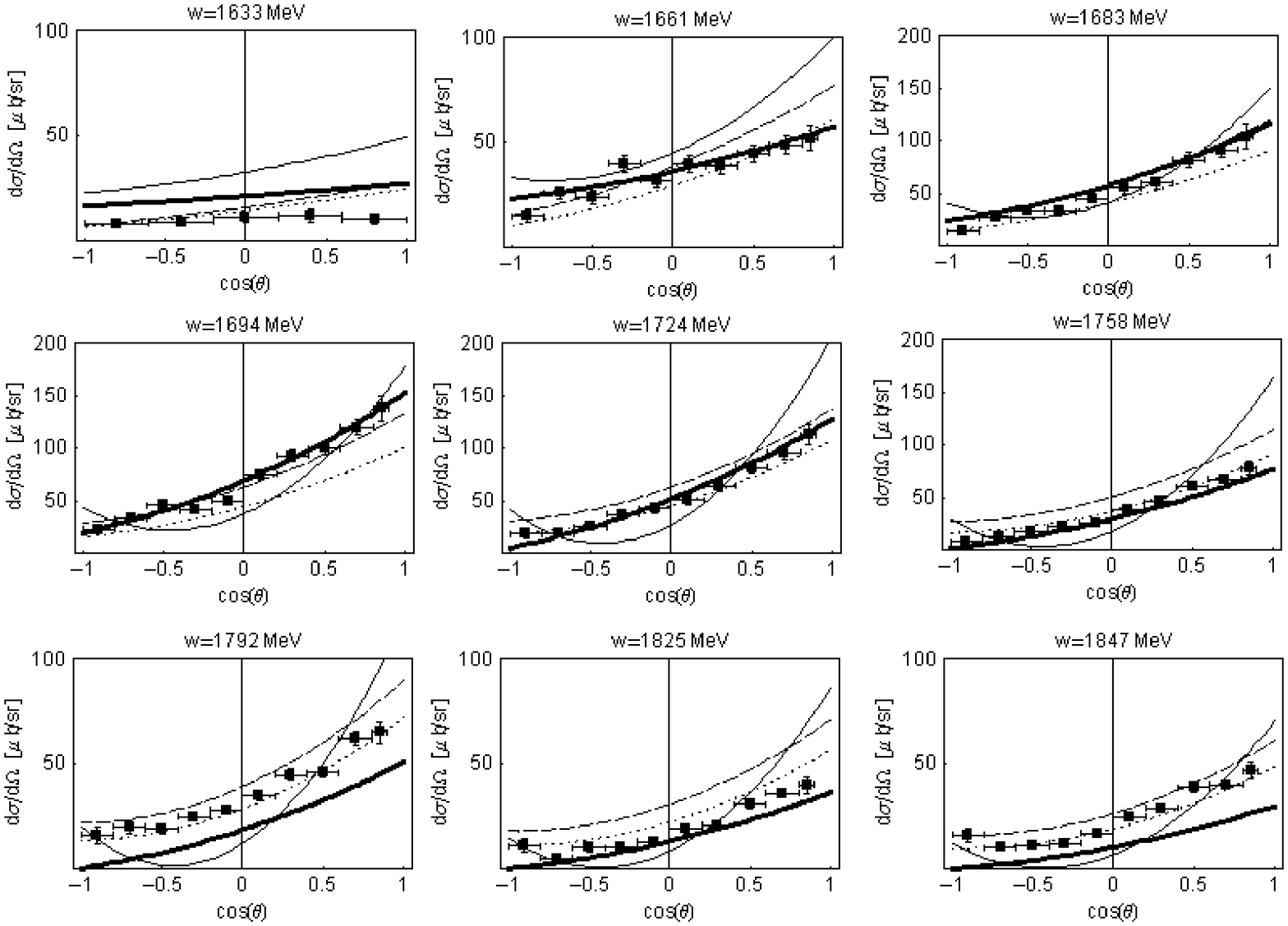,height=9.5cm}   
\caption{The agreement of the available experimental data for the differential cross section\protect\cite{Bak78} 
(full boxes) with the single resonance model predictions using different inputs for the resonance parameters: 
"standard" (PDG) set (thin solid line); $Sol \ 1$ (dotted line); $Sol \ 2$ (dashed line) and $Sol \ 3$ (thick solid 
line).
\label{inter}}
\end{center}
\end{figure}

   We fit the "lower"  and "upper" band  of the total cross section  imposing  the correct angular dependence at the 
same time, and obtain  $Sol \ 1$ (dotted line) and $Sol \ 2$ (dashed line).   To obtain the $Sol \ 3$ (thick solid 
line) we fit {\em only} Baker et al data\cite{Bak78}. \\ 
 "Standard" (PDG) solution introduces "too much curvature" in the differential cross section throughout the whole 
energy range indicating too big P-waves contribution relative to the S-wave.
  If the shape of the angular dependence is to be reproduced, contrary to the standard belief\cite{PWA,PDG}, the 
contribution of $P_{13}$ partial wave is negligible for all obtained solutions.
The branching ratio of $S_{11}$ resonance to $K \Lambda$ channel is somewhat smaller then previously believed.
The branching ratio of $P_{11}$ resonance to $K \Lambda$ channel is significantly bigger.
 If the latest Baker et al data\cite{Bak78} are to be taken very seriously the best agreement with the experiment is 
obtained for {\em very narrow} $P_{11}$ resonance, not observed in other processes and strongly inelastic - $Sol \ 
3$; hence a candidate for a non-strange pentaquark counter-partner.

The re-measuring of the differential cross section for the $\pi$N $\rightarrow$ K$ \Lambda$ process  in the energy 
range 1600 MeV $<$ w $<$ 1800 MeV is badly needed. The decisive conclusion about the existence of the $P_{11}$ 
non-strange pentaquark counter-partner will be possible only when the improved set of data is fully incorporated in 
one of the existing coupled channel partial wave \vspace*{-0.2cm} analyses\cite{PWA}.


\begin{thebibliography}{0}

\bibitem{Bak78} R.D. Baker et. al., {\it Nucl. Phys.} {\bf B141}, 29 (1978).
\bibitem{Dat78} Landolt-B\"{o}rnstein, New Series, ed. H. Schopper, {\bf 8} (1973). 
\bibitem{Arn85} R.A. Arndt, J.M. Ford and L.D. Roper, {\it Phys. Rev.} {\bf D32}, 1085 (1985).
\bibitem{Arn04}R. A. Arndt et. al., {\it Phys. Rev.} {\bf C69}, 035213 (2004).
\bibitem{PWA} R.E. Cutkosky, C.P. Forsyth, R.E. Hendrick and R.L. Kelly, {\it Phys. Rev.} {\bf D20}, 2839 (1979);
  M. Batini\'{c}, I. \v{S}laus, A. \v{S}varc and B.M.K. Nefkens, {\it Phys. Rev} {\bf C51}, 2310 (1995);
  T.P. Vrana, S.A. Dytman and T.S.-H- Lee, {\it Phys. Rep.} {\bf 328}, 181 (2000).
\bibitem{Pen04} T. Nakano et al., {\it Phys. Rev. Lett.} {\bf 91}, 012002 (2003); : NA49 Collaboration,
{\it Phys.Rev.Lett.} {\bf 92}, 042003 (2004).
\bibitem{Dia97} D. Diakonov, V. Petrov and M. Polyakov, {\em Z. Physik} {\bf A359}, 305 (1997).
\bibitem{Jaf03} R. Jaffe and F. Wilczek, {\em Phys. Rev. Lett.}  {\bf 91}, 232003 (2003).
\bibitem{Tsu00} K. Tsushima, A. Sibirtsev and A.W. Thomas, {\it Phys.Rev.} {\bf C62}, 064904  (2000).
\bibitem{PDG} K. Hagiwara et.al., {\it Phys. Rev.} {\bf D66}, 010001 (2002.)
\bibitem{Bat98} M. Batini\'{c}, I. Dadi\'{c}, I. \v{S}laus, A. \v{S}varc, B.M.K. Nefkens and T.S.-H. Lee, {\it 
Physica Scripta} {\bf 58}, 15 (1998)

\end{thebibliography}
\end{document}